\documentclass[letter]{IEEEtran}
\usepackage{amssymb}
\usepackage{amsmath}
\usepackage{amsfonts}
\usepackage{graphicx}
\usepackage{algorithm}
\usepackage{algorithmic}
\usepackage{epstopdf}
\usepackage{color}
\usepackage{multirow}
\usepackage{lettrine}
\usepackage{caption}
\usepackage{subcaption}
\usepackage{amsmath,amsfonts,amssymb,amsthm, bm}
\usepackage{xcolor,soul,framed}
\usepackage[noadjust]{cite}
\usepackage{filecontents}
\usepackage{multibib}
\usepackage{relsize}
\usepackage{lipsum}
\normalsize
\usepackage{multicol,lipsum,xparse}

\ifCLASSINFOpdf
\else
\fi
\begin{document}
%
\title{Subcarrier Assignment and Power Allocation for SCMA Energy Efficiency}

\author{Samira~Jaber, Wen~Chen, Kunlun Wang, and Jun Li
\thanks{S. Jaber and W. Chen are with Department of Electronic Engineering, Shanghai Jiao Tong University, China. (email:\{samira;wenchen\}@sjtu.edu.cn).}
\thanks{K. Wang is with the School of Information Science and Technology, ShanghaiTech Unitersity, China. (e-mail: fwangkl2@shanghaitech.edu.cn). }
\thanks{J. Li is with the School of Electronic and Optical Engineering, Nanjing University of Science Technology, China	(e-mail: jun.li@njust.edu.cn).}
}
%
\markboth{IEEE Journal}%
{Submitted paper}
\maketitle
\begin{abstract}
In this paper we propose resource allocation algorithm for uplink sparse code multiple access (SCMA) networks to maximize the energy efficiency (EE). Due to the joint optimization of factor graph  matrix and power allocation matrix, the EE maximization is a non-convex mixed-integer non-linear program (MINLP) problem. After transforming the non-convex form of the uplink sum rate to a convex one and separating subcarrier assignment and power allocation, we propose an energy efficient subcarrier assignment algorithm. By applying the fractional programming theory based on Dinkelbach method, we then propose power allocation algorithm to maximize EE. Finally, the simulation results show that the proposed resource allocation algorithm can significantly increase the EE of the uplink SCMA network.	
\end{abstract}
\section{Introduction}
Sparse Code Multiple Access (SCMA) \cite{wei2016scma}
illustrating a promising candidate for the 5G system, is a multi-dimensional codebook based non-orthogonal spreading technique that supplies a near-optimal design of sparse codewords and achieves considerable progress in spectral efficiency and massive connectivity.
Mobile communication networks contribute significantly towards the global carbon emission, thus, energy consumption is a critical concern when planning for 5G networks
\cite{wu2018noma,wu2014ofdma}. Therefore improving the energy efficiency (EE), which is defined as the achievable data rates over the total power consumption, has received much investigation recently.

In \cite{zhang2014sparse}, the EE of uplink SCMA network was analyzed and a comparison was made between the performance of average block error rate (BLER) and EE for the SCMA scheme and LTE-A. The obtained simulation results demonstrate that the SCMA scheme is capable of aggregating more users in the uplink network to support the massive connectivity requirements for 5G systems with reasonable energy consumption. In addition to considering equal transmit power of nonzero elements for all users, a fixed factor graph matrix is also used in this reference for problem analysis.
The codebook assignment and optimal power allocation for downlink SCMA is investigated in \cite{li2016cost}.
The obtained simulation results reveal that the SCMA network significantly outperforms the OFDMA network in term of EE. Moreover, in \cite{jin2018resource}, resource allocation for EE maximization of layered multicast in downlink SCMA network, by separating codebook assignment and power allocation was considered.


\cite{zhai2015rate} investigated the rate and energy maximization problem in SCMA networks with simultaneous wireless information and power transfer. In \cite{dong2016energy} cooperative co-evolutionary particle swarm optimization (CCPSO) was applied to solve the EE maximization problem for uplink SCMA network. In \cite{huang2018power} the Lagrange dual decomposition method and Dinkelbach theory to solve the non-convex optimization problem to maximize the EE for SCMA downlink systems is employed, based on which a three-level power allocation to improve the sum capacity for SCMA downlink system is proposed in \cite{han2018optimal}. The authors in \cite{d2018learning} proposed a computationally efficient algorithm to maximize the global energy efficiency with subject to both maximum power and minimum transmission rate constraints in the multi-carrier wireless interference network. Finally, it is converted into a non-convex fractional problem and was tackled through an interplay of fractional programming based on Dinkelbach algorithm, fixed-point learning and the generalized game theory. In \cite{yu2019power} the EE problem was approximately decomposed  into some independent convex sub-problems by applying the fractional programming theory and using the coordinate descent method, and the closed-form solutions of each sub-problem are derived.

\cite{dong2016energy} evaluates EE of the network using the power allocation matrix without considering different path-loss. On the other hand, solving the resource allocation problem using separate subcarrier assignment and power allocation \cite{li2015energy} is useful in analyzing downlink SCMA EE, especially in reducing the computational cost \cite{li2016cost}. Moreover, \cite{huang2018power} employs the water-filling algorithm to solve the non-convex power allocation optimization problem to evaluate the performance of downlink SCMA in term of EE with fixed factor graph. Briefly, we need to develop optimal resource allocation algorithms for uplink SCMA scheme by incorporating the factor graph matrix optimization to increase energy efficiency that meets 5G network requirements.

The aim of this paper is to maximize the EE for an uplink SCMA system where the the path-losses between each user and the base station (BS) are not necessarily equal. 
The EE optimization problem involving subcarrier assignment and power allocation is a non-convex mixed-integer non-linear program (MINLP) problem~\cite{ghasemishabankareh2019genetic}. To reduce the computational cost, we separate subcarrier assignment and  power allocation. A fast subcarrier assignment algorithm with power equally distributed is proposed to determine an energy efficient optimal and of low-complexity factor graph matrix. For specified factor graph matrix, a power allocation algorithm is proposed for EE maximization, which is decomposed into convex sub-problems by applying the fractional programming theory based on Dinkelbach method.

\section{System model and problem formulation}

\subsection{System model}
We consider an SCMA based uplink communication network that is comprised of one based station (BS) and $J$ users, where an individual user represents a layer of SCMA. BS and all the users are equipped with a single antenna. There are totally $K$ subcarriers available where the $J$ users are multiplexed over these subcarriers. The number of codewords is equal to the number of users and the length of each codeword is $K$. The transmitted codewords for the $j$th user is represented by $x_{j}=\left [x_{j,1},...,x_{j,K} \right ]^{T}$, which has $N$ nonzero elements. The zero and nonzero entries for all transmitted signals are represented by the $K\times J$ factor graph matrix $F$. According to the overloading feature of SCMA scheme, we have $J> K$ and $J/K$ is called overloading factor.

Let $f_{j}=\left [f_{j,1},...,f_{j,K} \right ]^{T}$ denote the $j$th column vector of matrix $F$, called indicator vector of the $j$th user, where $f_{j,k}=1$ if $x_{j,k}\neq 0$, and $f_{j,k}=0$ otherwise. Let $p_{j,k}$ be the power of the $j$th user in the $k$th subcarrier, where $p_{j,k}\neq 0$ if $f_{j,k}=1$, and $p_{j,k}= 0$, if $f_{j,k}=0$.
Let $h_{j}=[ h_{j,1},...,h_{j,K}]^{T}$ be the channel vector of the $j$th user, and $h_{j,k}$ is the channel gain of the $j$th user in the $k$th subcarrier. Let $n$ be the additive white Gaussian noise. Then the received signal  $y=[y_{1},...,y_{K}]^{T}$ in BS, which is a combination of all propagated signals through $K$ subcarriers, is written as follows.
\begin{equation}\label{e1}
y=\sum_{j=1}^{J}diag\left (p_{j}  \right )diag\left ( f_{j} \right )diag\left ( h_{j} \right )x_{j}+n,
\end{equation}
where $p_{j}= [\sqrt{p_{j,1}},...,\sqrt{p_{j,k}}]^{T}$. Using the sparsity in the SCMA scheme, the message passing algorithm (MPA) detector can be applied to achieve a near-optimal detection performance. With the knowledge of the channel state information (CSI), the data rate of the $j$th user in the $k$th subcarrier is \cite{d2018learning},
\begin{equation}\label{e2}
R_{j,k}=\log_{2}\left [1+ \frac{f_{j,k}p_{j,k}\left |h_{j,k} \right |^{2}}{\sigma ^{2}+\sum_{t=1,t\neq j}^{J}f_{t,k}p_{t,k}\left |h_{t,k}  \right |^{2}} \right ].
\end{equation}
where $\sigma^{2}$ is the noise power.
The achievable rate in the $k$th subcarrier is the sum of all users rates assigned to this subcarrier,
\begin{equation}\label{e3}
R_{k}=\sum_{j=1}^{J}R_{j,k}.
\end{equation}
Using (\ref{e2}) and (\ref{e3}), the sum rate of the network is
\begin{equation}\label{e4}
R=\sum_{k=1}^{K}R_{k}=\sum_{k=1}^{K}\sum_{j=1}^{J}R_{j,k}.
\end{equation}
According to \cite{li2015energy,li2016cost,dong2016energy}, the total power consumption of the whole system is
\begin{equation}\label{e5}
P_{tot}=\sum_{j=1}^{J}\sum_{k=1}^{K}f_{j,k}p_{j,k}+J\times P_{c}.
\end{equation}
where $P_{c}\geq 0$ is the fixed circuit power consumption of each user. In (\ref{e5}), the total power consumption for the uplink is linearly increasing with $p_{j,k}$, and obviously, which has a critical role on the total power consumption of the system. As per \cite{li2015energy,li2016cost,dong2016energy}, EE is the total  uplink sum rate of all users over the total power consumption. Thus, from (\ref{e4}) and (\ref{e5}), the energy efficiency $\eta_{EE}$ is
\begin{equation}
\eta_{EE}\overset\triangle =\frac{R}{P_{tot}}=\frac{R\left ( F,P \right )}{P_{tot}\left ( F,P \right )},
\end{equation}
where ${P=\{p_{j,k}\}}$ is the power allocation matrix of all users of the network.

\subsection{Problem formulation}
As mentioned above, it is crucial to obtain the factor graph matrix $F$ and transmit power allocation matrix $P$ to improve the EE of the network. This is more important when SNRs are different for network users and users are distributed in different locations within the network. The EE maximization problem can be formulated as
\begin{equation}\label{e7}
\begin{split}
&{\max_{F,P}} \eta_{EE}\left ( F,P \right ),\\
s.t.\,\,&C_{1}:\sum_{k=1}^{K}f_{j,k}=N, \, 1\leq j\leq J,\\
&C_{2}:f_{j,k}\in \left \{ 0,1 \right \},  1\leq j\leq J, \, 1\leq k\leq K,\\
&C_{3}:f_{j,}\neq f_{{j}'}, \forall j, {j}'\in J, j\neq {j}',\\
&C_{4}:\sum_{k=1}^{K}p_{j,k}\leq p_{j}^{\max}, \, 1\leq j\leq J,\\
&C_{5}:p_{j,k}\geq 0, \, 1\leq k\leq K,
\end{split}
\end{equation}
where $P_j^{\max}$ is the maximum power of user $j$. The constraint $C_{1}$ shows the number of subcarriers that each user occupies, and according to the conditions of the SCMA network, $N$ is the number of nonzero symbols of users. The constraint $C_{2}$ shows the feature of the factor graph matrix. The constraint $C_{3}$ denotes that $F$ have no two same columns. There are $\binom{K}{N}$ possible states for the first column, and there is one state decrement for the subsequent columns. Hence, the number of all different possible $F$ is
\begin{equation}\label{e8}
\prod_{j=1}^{J}\left(\binom{K}{N}-\left ( j-1 \right ) \right).
\end{equation}
The constraint $C_{4}$ shows the limitation of each user's transmit power, and $C_{5}$ shows that each power allocation coefficient must be non-negative.
Optimization problem (\ref{e7}) is a non-convex mixed-integer non linear programming (non-convex MINLP) problems~\cite{ghasemishabankareh2019genetic}.

\section{Fast subcarriers assignment and power allocation}
Considering the received signal in each subcarrier, the sum capacity of uplink network can be rewritten as
\begin{align}\label{e9}
\begin{aligned}
R\doteq \sum_{k=1}^{K}\log_{2}\left ( 1+\frac{\sum_{j=1}^{J}f_{j,k}p_{j,k}\left |h_{j,k}  \right |^{2}}{\sigma ^{2}} \right ).
\end{aligned}
\end{align}
Eq. (\ref{e9}) implies that both $f_{j,k}$ and $p_{j,k}$ have important roles on the SCMA sum rate. The problem (\ref{e7}) can be written as

\begin{align}\label{e10}
\begin{aligned}
&\underset{F,P}{\max}\frac{\sum_{k=1}^{K}\log_{2}\left ( 1+\frac{\sum_{j=1}^{J}f_{j,k}p_{j,k}\left |h_{j,k} \right |^{2}}{\sigma^{2}} \right )}{\sum_{j=1}^{J}\sum_{k=1}^{K}f_{j,k}p_{j,k}+J\times P_{c}},\\
s.t.\,\, &C1, C2, C3, C4, C5.
\end{aligned}
\end{align}

\subsection{Fast Subcarrier Assignment Algorithm}
This algorithm is to allocate the subcarrier to users one by one. At each stage, calculate the network EE increment. When the $j$th user access to the network, the $j$-th column $f_j$ of $F$ is updated to form the new factor graph $F(j)$. The EE increment is calculated as
\begin{equation}
\Delta\eta_{EE}(F(j),P)=\eta_{EE}(F(j),P)-\eta_{EE}(F(j-1),P).
\end{equation}

Determine the optimal $F$ under the condition that the transmit power is equally distributed for user symbols, meaning that power is equally distributed among codebooks. Generate a factor graph matrix $S$ randomly. When $jN\leq K$, $f_j$ can be obtained by allocating orthogonal column of $S$. When $jN> K$,
the $\Delta \eta _{EE}(F(j),P)$ is calculated as follows.
\begin{multline}
\Delta \eta _{EE}(F(j),P)=\frac{\sum_{k=1}^{K}\log_{2}\left ( 1+\frac{\sum_{t=1}^{j}f_{t,k}p_{t,k}\left | h_{t,k} \right |^{2}}{\sigma^{2} } \right )}{\sum_{t=1}^{j}\sum_{k=1}^{K}f_{t,k}p_{t,k}+t\times P_{c}}\\
-\frac{\sum_{k=1}^{K}\log_{2}\left ( 1+\frac{\sum_{t=1}^{j-1}f_{t,k}p_{j,k}\left | h_{t,k} \right |^{2}}{\sigma ^{2}} \right )}{\sum_{t=1}^{j-1}\sum_{k=1}^{K}f_{t,k}p_{t,k}+\left ( t-1 \right )\times P_{c}}.
\end{multline}
Then $f_j$ can be obtained by solving the following optimization problem over the rest columns of $S$, i.e.,
\begin{equation}\label{e13}
\begin{split}
&\max_{S} \Delta \eta_{EE}(F(j),P),\\
s.t.\,\, & C1,\, C2,\, C3.
\end{split}
\end{equation}
This algorithm is
summarized in the following Algorithm~$1$.
\begin{algorithm}[H]
	\textbf{Input:} $K$, $J$, $P_{\max}$.\\
	\textbf{Initialization:} Generate a $K\times J$ matrix $S$ randomly. Set $j=0$, $F=\left [ 0 \right ]_{K\times J}$, $p_{j}^{\max}=P_{\max}$, and $p_{j,k}=\frac{P_{max}}{N}$;\\
	\textbf{While} $j\leq J$ \textbf{do}\\
	Update  $j=j+1$;\\
	\textbf{If} $jN\leq K $ \textbf{then}\\
	Select $f_{j}\in S$ with $\left \{ f_{j} \right \}\cap F=\phi$;\\
		Update $F=F\cup \left \{ f_{j}\right \}$, $S=S\setminus \left \{ f_{j} \right \}$;\\
		\textbf{end if}\\
		Find one solution $f_{j}$ of problem (\ref{e13});\\ 
		Update $F=F\cup \left \{f_{j} \right \}$, $S=S\setminus \left \{ f_{j} \right \}$;\\
	\textbf{end while}\\
	\textbf{Output} the optimal factor graph matrix: $F=F(J)$.
	\caption{Fast subcarrier assignment algorithm}
\end{algorithm}

\subsection{Energy efficient power allocation}
There is a tradeoff between the sum rate of the network and the total power consumption for the EE maximization problem while users have limitation on the transmit power.
The optimization problem (\ref{e10}) is a non-convex problem due to the fractional-form of the objective function. The fractional programming based on Dinklebach's algorithm \cite{dinkelbach1967nonlinear} can be exploited for solving this problem~\cite{d2018learning}.
It is observed that the numerator and denominator of the objective function in (\ref{e10}) are both concave. After $F$ is determined according to subcarrier algorithm in the previous section. The optimization problem  (\ref{e10}) can be written as
\begin{equation}\label{e14}
\begin{split}
&\max_P\eta _{EE}\left ( P \right ),\\
s.t.\,\, &C4, C5.
\end{split}
\end{equation}
Define an auxiliary function
\begin{multline}\label{e15}
A(\omega^{(t)},P^{(t)})\overset{\triangle}=\sum_{k=1}^{K}\log_{2}\left ( 1+\frac{\sum_{j=1}^{J}f_{j,k}p_{j,k}^{(t)}\left | h_{j,k} \right |^{2}}{\sigma ^{2}} \right )\\
-w^{\left (t  \right )}\left ( \sum_{k=1}^{K}\sum_{j=1}^{J}f_{j,k}p_{j,k}^{(t)} +J\times P_{c}\right),
\end{multline}
where $w^{\left (t  \right )}$ is a non-negative factor. Based on Dinklebach method, we can transform problem (\ref{e14}) into a series of parametric subtractive-form sub-problems as given as follows~\cite{yu2019power}
\begin{align}\label{e16}
\begin{aligned}
& \max_{P^{(t)}}  A(\omega^{(t)},P^{(t)}),\\
s.t.\,\, &C4, C5.
\end{aligned}
\end{align}

According to Dinklebach algorithm, the value of $\omega^{\left (t  \right )}$ starts from $0$ and it is updated to its optimal value by using an iterative procedure. The procedure is terminated when the value of auxiliary function $A(w^{\left (t  \right )},P^{(t)})$ is smaller than threshold $\varepsilon$, i.e., $A(w^{\left (t  \right )},P^{(t)})< \varepsilon $. As $t$ increases, the value of $A(w^{\left (t  \right )},P^{(t)})$ decreases. The term $p_{j,k}^{(t)}$ in $A(w^{\left (t  \right )},P^{(t)})$ denotes the updated power after solving (\ref{e16}). Once the iterative procedure is terminated, the maximum value of $\eta _{EE}$ is $w^{\left (t  \right )}$.

Since $A\left (w^{(t)},P^{(t)} \right )$ is a concave function with respect to $p_{j,k}^{(t)}$, we can use Lagrangian multiplier method to solve problem~(\ref{e16}). The Lagrangian function is defined as
\begin{align}
\begin{aligned}
L\left ( \lambda^{(t)} ,P^{(t)} \right )&=\sum_{k=1}^{K}\log_{2}\left ( 1+\frac{\sum_{j=1}^{J}f_{j,k}p_{j,k}^{(t)}\left | h_{j,k} \right |^{2}}{\sigma ^{2}} \right )\\
&-\omega^{\left ( t \right )}\left ( \sum_{j=1}^{J}\sum_{k=1}^{K}f_{j,k} p_{j,k}^{(t)}+\left ( J\times P_{c} \right )\right )\\
&-\sum_{j=1}^{J}\lambda _{j}^{(t)}\left ( \sum_{k=1}^{K} f_{j,k}p_{j,k}^{(t)}-p_{j}^{\max}\right ),
\end{aligned}
\end{align}
where $\lambda^{(t)}=\{\lambda^{(t)}_j,\dots,\lambda^{(t)}_J\}$ denotes the Lagrange multiplier. In addition, in terms of the determined $F$, we have $\sum_{j=1}^{J}\sum_{k=1}^{K}f_{j,k}p_{j,k}=\sum_{k=1}^{K}\sum_{j=1}^{J}f_{j,k}p_{j,k}$. Then the derived transmit power satisfying Karush-Kuhn-Tucker conditions~\cite{boyd2004convex} is
\begin{multline}\label{e18}
p_{j,k}^{(t+1)}=\max \left[0,\frac{1}{\left ( \lambda _{j}^{(t)} +\omega^{\left ( t \right )}\right )\ln{2}}\right.\\
\left.-\frac{\sum_{j=1}^{J}f_{j,k}p_{j,k}^{(t)}\left | h_{j,k} \right |^{2}+\sigma ^{2}}{f_{j,k}\left | h_{j,k} \right |^{2}} \right].
\end{multline}
Once $p^{(t+1)}_{j,k}$ is derived, the Lagrangian multiplier can be updated using the following subgradient algorithm~\cite{huang2018power},
\begin{align}\label{e19}
\begin{aligned}
\lambda _{j}^{(t+1)}=\max\left [0,\lambda _{j}^{\left ( t \right )}-\beta_j\left ( p_{j}^{\max} -\sum_{k=1}^{k}p_{j,k}^{\left ( t +1\right )}\right )\right ],
\end{aligned}
\end{align}
where $\beta_j$ is the iteration step size. Subsequently, after a limited number of iterations, (\ref{e19}) will converge. Additionally, to prevent the reduction in the sum rate of the SCMA network (\ref{e9}), it can be assumed that each user utilizes the maximum transmit power. In other words, the sum of transmit power of non-zero elements of each user is $p_{j}^{\max}$. To prevent the reduction in the sum rate of the network in addition to optimizing EE, we can address the calculation of (\ref{e18}) via initialization of the uplink network power into a feasible values, while the Lagrange multiplier satisfies the equation $\sum_{k=1}^{K}p^{(t)}_{j,k}=p_{j}^{\max}$. Having the derived $F$, subcarrier assignment to non-zero elements of users is determined. Therefore, it is merely required to design $\lambda_{j}^{(t)}$ in order to satisfy $\sum_{k=1}^{K}{p}_{j,k}^{(t)}=p_{j}^{\max}$ for $J$ different users. The energy efficient power allocation algorithm is summarized in the following Algorithm~$2$.
\begin{algorithm}[H]
	\textbf{Input:} $\textbf{F}$ $\leftarrow$ the solution of Algorithm $1$ .\\
	\textbf{Initialization:} $t=1$, $\varepsilon=10^{-6}$, $\omega^{\left (1  \right )}=0$, $A(\omega^{\left (1  \right )},P^{(1)})=1$, $\lambda_{j}^{(1)}=1$, $\beta_j=1$.\\
	\textbf{While} $A\left (\omega^{\left (t  \right )} , P^{(t)} \right )\geq \varepsilon $ \textbf{do}\\
	Update $p^{(t)}_{k,j}$ by (\ref{e18});\\
	Update $\lambda_j^{(t)}$ by (\ref{e19});\\
	Update $\omega^{\left ( t \right )}=\frac{\sum_{k=1}^{K}\log_{2}\left ( 1+\frac{\sum_{j=1}^{J}f_{j,k}p_{j,k}^{\left ( t \right )}\left | h_{j,k} \right |^{2}}{\sigma ^{2}} \right )}{\sum_{j=1}^{J}\sum_{k=1}^{K}f_{j,k}p_{j,k}^{\left ( t \right )}+ \left (J\times P_{c}  \right )}$;\\
	Update $A(\omega^{(t)},P^{(t)})$ by (\ref{e15});\\
	Update $t=t+1$;\\
	\textbf{end} while\\
	\textbf{Output} $\eta _{EE}(F,P)=\omega^{\left ( t \right )}$.
	\caption{Energy efficient power allocation algorithm}
\end{algorithm}

\section{Simulation results}
We consider a single-cell uplink network where $J=6$ users are randomly distributed in a circular area of $100$m radius. In the simulation, both small-scale fading and large-scale path loss have been considered. The channel gain $h_{j,k}=g_{j,k}d_{j}^{-\frac{\alpha }{2}}$ where $g_{j,k}$ represents the Rayleigh fading channel gain of the $j$th user on the $k$th subcarrier, and $\alpha =3.67$ is the path loss exponent~\cite{yang2017uplink}. The numbers of  subcarriers and non-zeros elements in each SCMA-codeword are $K=4$ and $N=2$, respectively. The bandwidth of each subcarrier, the power density of noise and the fixed circuit power consumption at each user are $180kHz$, $-174dBm/Hz$ and $1mW$, respectively.

The simulation results are averaged over $150$ experiments. The computational complexity of Algorithm~$1$ is $\left ( J-\left \lfloor \frac{K}{N} \right \rfloor \right )!$. In the case $J=6$, $K=4$, and $N=2$, the number of iterations is $\left ( 6-\left \lfloor \frac{4}{2} \right \rfloor \right )!=24$, while, according to (\ref{e8}), the number of iterations is $720$ for the exhaustive search.

Fig. 1 and Fig. 2, present the energy efficient SCMA uplink system for four different cases of resource allocation defined as follows.
\begin{itemize}
	\item \textbf{SCMA-PA-PPC:} The subcarrier assignment uses the proposed assignment (PA) Algorithm~$1$, and the power allocation uses the proposed power allocation Algorithm~$2$ under the power constraints (PPC): $\sum_{k=1}^{K}p_{j,k}\leq p_{j}^{\max}$.
	\item \textbf{SCMA-PA-PMP:} The subcarrier assignment uses the proposed assignment (PA) Algorithm~$1$, but  the power allocation uses the proposed power allocation Algorithm~$2$ under the maximum power constraint (PMP): $\sum_{k=1}^{K}p_{j,k}= p_{j}^{\max}$.
	\item \textbf{SCMA-RA-PMP:} Random assignment (RA) method refers to the method that randomly assigns subcarriers to users \cite{li2016cost,xiong2018optimal}. The power allocation uses Algorithm~$2$ under the maximum power constraint (PMP).
	\item \textbf{SCMA-FA-PMP:} In fixed subcarrier assignment (FA) method we use a factor graph matrix that it’s given by \cite{zhang2014sparse} and \cite{yang2017uplink}, and the power allocation uses Algorithm~$2$ under the maximum power constraint (PMP).	
\end{itemize}

Fig.~1 shows the EE for the four different subcarrier assignments and power allocations, where all the user distances are equal to $100$m. It is shown that SCMA-PA-PPC and SCMA-PA-PMP outperforms SCMA-RA-PMP and SCMA-FA-PMP. Therefore the proposed subcarrier assignment performs better. Moreover both SCMA-PA-PPC and SCMA-PA-PMP converge to a similar maximum level as $P_{\max}$ increases. It is also shown that SCMA-PA-PPC has a better performance compared to SCMA-PA-PMP. This is because that each user using the maximum power has more fairness but loses EE performance.
\begin{figure}[H]
	\centering
	\includegraphics[width=3.5in]{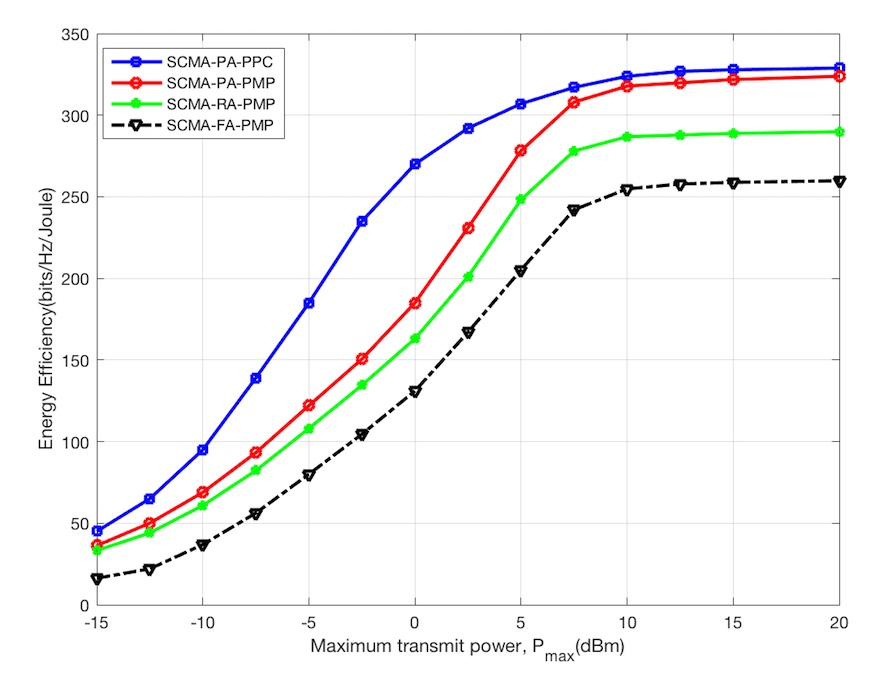}
	\caption{Energy efficiency for different cases of resource allocation where all user distances are equal to $100$m.}
	\vspace{-0.3em}
	\label{SCMAA}
\end{figure}
Fig. 2 considers the influence of user distances to the uplink SCMA EE. Two different conditions are considered for users with the same average location, which are listed in the following table.
\begin{table}[H]
	  \centering
	  \caption{User distance condition}
	\begin{tabular}{|c|c|c|}
		\hline
		\textbf{User distance}   & \textbf{Condition 1} & \textbf{Condition 2}      \\ \hline
		User~$1$ distance        &$55$m                   & $77$m                    \\ \hline
		User~$2$ distance        &$68$m                   & $80$m                     \\ \hline
		User~$3$ distance        &$89$m                   & $81$m                     \\ \hline
		User~$4$ distance        &$99$m                   & $90$m                     \\ \hline
		User~$5$ distance        &$99$m                   & $91$m                     \\ \hline
		User~$6$ distance        &$100$m                  & $91$m                     \\ \hline
		Average user distance    &$85$m                   & $85$m                     \\ \hline
	\end{tabular}
\end{table}

From Fig.~2, it is observed that SCMA-PA-PPC has almost the same performance for the user distance condition 1 and condition 2. So does SCMA-PA-PMP.
This is due to the proposed method which offers the optimal factor graph that leads to the maximum EE via applying an equal maximum transmit power to each non-zeros element of all users. It is clear that there is a significant increase in EE for SCMA-PA-PPC and SCMA-PA-PMP over SCMA-RA-PMP and SCMA-FA-PMP. The EE for both SCMA-RA-PMP and SCMA-FA-PMP has better performance for the user distance condition 1 than that for condition 2.

From Fig. 2, it can be seen, the EE performance for condition 1 is better than that for condition 2, which implies that the EE performance is mainly influenced by the users close to the base station. These users close to base station are more likely to have the most important contribution to optimize the energy use.

\begin{figure}[H]
	\centering
	\includegraphics[width=3.5in]{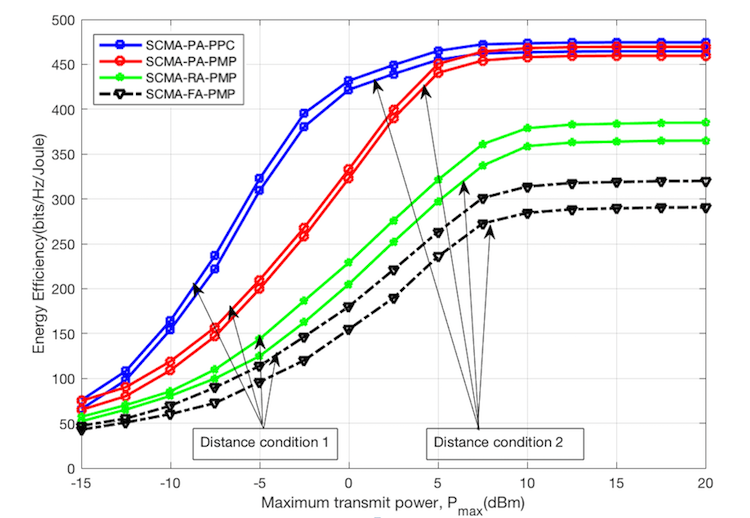}
	\caption{The influence of user distances with the same average location on the energy efficiency for different cases of resource allocation.}
	\vspace{-0.3em}
	\label{SCMADIS}
\end{figure}

\section{Conclusion}
In this letter, we proposed a fast subcarrier assignment algorithm for energy efficient uplink SCMA. Then for a determined SCMA factor graph matrix, we propose a power allocation algorithm to solve the non-convex optimization problem using Dinklebach method. By the proposed algorithms, the SCMA EE is well optimized based on subcarrier assignment and power allocation, while the complexity of Algorithm~$1$ is much decreased compared to the exhaustive search. Simulations show that the proposed algorithm outperforms the random and fixed factor graph matrix, and users closed to base station contribute more to the SCMA EE.
\bibliographystyle{ieeetran}
\bibliography{Ref}

\begin{thebibliography}{10}
\providecommand{\url}[1]{#1}
\csname url@samestyle\endcsname
\providecommand{\newblock}{\relax}
\providecommand{\bibinfo}[2]{#2}
\providecommand{\BIBentrySTDinterwordspacing}{\spaceskip=0pt\relax}
\providecommand{\BIBentryALTinterwordstretchfactor}{4}
\providecommand{\BIBentryALTinterwordspacing}{\spaceskip=\fontdimen2\font plus
\BIBentryALTinterwordstretchfactor\fontdimen3\font minus
  \fontdimen4\font\relax}
\providecommand{\BIBforeignlanguage}[2]{{%
\expandafter\ifx\csname l@#1\endcsname\relax
\typeout{** WARNING: IEEEtran.bst: No hyphenation pattern has been}%
\typeout{** loaded for the language `#1'. Using the pattern for}%
\typeout{** the default language instead.}%
\else
\language=\csname l@#1\endcsname
\fi
#2}}
\providecommand{\BIBdecl}{\relax}
\BIBdecl

\bibitem{wei2016scma}
F.~Wei and W.~Chen, ``Low complexity iterative receiver design for sparse code
  multiple access,'' \emph{IEEE Transactions on Communications}, vol.~65,
  no.~2, pp. 621--634, 2016.

\bibitem{wu2018noma}
Q.~Wu, W.~Chen, D.~Ng, and R.~Schober, ``Spectral and energy-efficient wireless
  powered iot networks: Noma or tdma?'' \emph{IEEE Transactions on Vehicular
  Technology}, vol.~67, no.~7, pp. 6663--6667, 2018.

\bibitem{wu2014ofdma}
Q.~Wu, W.~Chen, M.~Tao, J.~Li, H.~Tang, and J.~Wu, ``Resource allocation for
  joint transmitter and receiver energy efficiency maximization in downlink
  ofdma systems,'' \emph{IEEE Transactions on Communications}, vol.~63, no.~2,
  pp. 416--430, 2014.

\bibitem{zhang2014sparse}
S.~Zhang, X.~Xu, L.~Lu, Y.~Wu, G.~He, and Y.~Chen, ``Sparse code multiple
  access: An energy efficient uplink approach for 5g wireless systems,'' in
  \emph{2014 IEEE Global Communications Conference}.\hskip 1em plus 0.5em minus
  0.4em\relax IEEE, 2014, pp. 4782--4787.

\bibitem{li2016cost}
Y.~Li, M.~Sheng, Z.~Sun, Y.~Sun, L.~Liu, D.~Zhai, and J.~Li, ``Cost-efficient
  codebook assignment and power allocation for energy efficiency maximization
  in scma networks,'' in \emph{2016 IEEE 84th Vehicular Technology Conference
  (VTC-Fall)}.\hskip 1em plus 0.5em minus 0.4em\relax IEEE, 2016, pp. 1--5.

\bibitem{jin2018resource}
X.~Jin, X.~Wang, and D.~Wang, ``Resource allocation for energy efficiency
  maximization of layered multicast in scma networks,'' in \emph{2018 24th
  Asia-Pacific Conference on Communications (APCC)}.\hskip 1em plus 0.5em minus
  0.4em\relax IEEE, 2018, pp. 459--464.

\bibitem{zhai2015rate}
D.~Zhai, M.~Sheng, X.~Wang, Y.~Li, J.~Song, and J.~Li, ``Rate and energy
  maximization in scma networks with wireless information and power transfer,''
  \emph{IEEE Communications Letters}, vol.~20, no.~2, pp. 360--363, 2015.

\bibitem{dong2016energy}
Y.~Dong, L.~Qiu, and X.~Liang, ``Energy efficiency maximization for uplink scma
  system using ccpso,'' in \emph{2016 IEEE Globecom Workshops (GC
  Wkshps)}.\hskip 1em plus 0.5em minus 0.4em\relax IEEE, 2016, pp. 1--5.

\bibitem{huang2018power}
Y.~Huang, S.~Han, S.~Guo, W.~Meng, and C.~Li, ``Power allocation for scma
  downlink systems based on maximum energy efficiency,'' in \emph{2018 14th
  International Wireless Communications \& Mobile Computing Conference
  (IWCMC)}.\hskip 1em plus 0.5em minus 0.4em\relax IEEE, 2018, pp. 1197--1202.

\bibitem{han2018optimal}
S.~Han, Y.~Huang, W.~Meng, C.~Li, N.~Xu, and D.~Chen, ``Optimal power
  allocation for scma downlink systems based on maximum capacity,'' \emph{IEEE
  Transactions on Communications}, vol.~67, no.~2, pp. 1480--1489, 2018.

\bibitem{d2018learning}
S.~D’Oro, A.~Zappone, S.~Palazzo, and M.~Lops, ``A learning approach for
  low-complexity optimization of energy efficiency in multicarrier wireless
  networks,'' \emph{IEEE Transactions on Wireless Communications}, vol.~17,
  no.~5, pp. 3226--3241, 2018.

\bibitem{yu2019power}
X.~Yu, F.~Xu, K.~Yu, and X.~Dang, ``Power allocation for energy efficiency
  optimization in multi-user mmwave-noma system with hybrid precoding,''
  \emph{IEEE Access}, vol.~7, pp. 109\,083--109\,093, 2019.

\bibitem{li2015energy}
Y.~Li, M.~Sheng, C.~W. Tan, Y.~Zhang, Y.~Sun, X.~Wang, Y.~Shi, and J.~Li,
  ``Energy-efficient subcarrier assignment and power allocation in ofdma
  systems with max-min fairness guarantees,'' \emph{IEEE Transactions on
  Communications}, vol.~63, no.~9, pp. 3183--3195, 2015.

\bibitem{ghasemishabankareh2019genetic}
B.~Ghasemishabankareh, M.~Ozlen, X.~Li, and K.~Deb, ``A genetic algorithm with
  local search for solving single-source single-sink nonlinear non-convex
  minimum cost flow problems,'' \emph{Soft Computing}, pp. 1--17, 2019.

\bibitem{dinkelbach1967nonlinear}
W.~Dinkelbach, ``On nonlinear fractional programming,'' \emph{Management
  science}, vol.~13, no.~7, pp. 492--498, 1967.

\bibitem{boyd2004convex}
S.~Boyd and L.~Vandenberghe, \emph{Convex optimization}.\hskip 1em plus 0.5em
  minus 0.4em\relax Cambridge university press, 2004.

\bibitem{yang2017uplink}
Z.~Yang, X.~Lei, Z.~Ding, P.~Fan, and G.~K. Karagiannidis, ``On the uplink sum
  rate of scma system with randomly deployed users,'' \emph{IEEE Wireless
  Communications Letters}, vol.~6, no.~3, pp. 338--341, 2017.

\bibitem{xiong2018optimal}
G.~Xiong and J.~Sun, ``An optimal resource allocation algorithm based on sum
  rate maximization for uplink scma system,'' in \emph{2018 IEEE 18th
  International Conference on Communication Technology (ICCT)}.\hskip 1em plus
  0.5em minus 0.4em\relax IEEE, 2018, pp. 805--810.

\end{thebibliography}
\end{document}